\documentclass[twocolumn,showpacs,preprintnumbers,amsmath]{revtex4}
\usepackage{graphicx}
\usepackage{dcolumn}
\usepackage{bm}

\begin{document}

\title{Berry Phase and Fidelity in the Dicke model with $A^{2}$ term}
\author{Yu-Yu Zhang$^{1,2}$, Tao Liu$^{3}$, Qing-Hu Chen$^{2,1,\dag}$, Kelin
Wang$ ^{3,4}$}
\address{$^{1}$ Department of Physics, Zhejiang University,
 Hangzhou 310027, P. R. China.\\
$^{2}$ Center for Statistical and Theoretical Condensed Matter
Physics, Zhejiang Normal University, Jinhua 321004, P. R. China
\\
$^{3}$ Department of Physics, Southwest University of Science and
Technology, Mianyang 621010, P. R. China.\\
$^{4}$ Department of Mordern Physics, University of Science and
Technology of China, Hefei 230026, P. R. China.}

\draft
\date{\today}

\begin{abstract}
The instability, so-called the quantum-phase-like transition, in the
Dicke model  with a rotating-wave approximation for finite $N$ atoms
is investigated  in terms of the Berry phase and the fidelity. It
can be marked by the discontinuous behavior of these quantities as a
function of the atom-field coupling parameter. Involving an
additional field $A^{2}$ term, it is observed that  the instability
is not eliminated beyond the  characteristic atom-field coupling
parameter  even for  strong interaction of the bosonic fields,
contrarily to the previous studies.

\end{abstract}


\maketitle

\section{Introduction}

The Dicke model (DM)\cite{dicke} describes an ensemble artificial
two-level atoms coupling with a cavity device. It has been attracted
considerable attentions recently, mainly due to the fact that the
Dicke model is closely related to many recent interesting fields in
quantum optics and condensed matter physics, such as the
superradiant behavior by an ensemble of quantum dots
\cite{Scheibner} and Bose-Einstein condensates \cite {Schneble},
coupled arrays of optical cavities used to simulate and study the
behavior of strongly correlated systems\cite{Hartmann}, and
superconducting charge qubits\cite{Nori,AMZ}. It is known from the
previous studies\cite{Lieb,Emary,Chen} that  the full DM undergoes
the second-order quantum phase transition \cite{Sach}.

As claimed in Ref. \cite{Nar}, a sequence of instabilities,
so-called quantum-phase-like transitions, is involved in the problem
of an ensemble of two-level atoms system interacting with a bosonic
field in the rotating-wave approximation (RWA)\cite{BVT,BVT1,Buzek}.
As addressed Ref. \cite{KRZ}, the absence of field $A^{2}$ term from
the minimal coupling Hamiltonian in the approximation of the DM
leads to the possibility of the instability. In the presence of
$A^{2}$ term, the classical thermodynamic properties have been
studied previously \cite{KRZ1,KRZ2,KRZ3}. Whether the instabilities
disappear when the interaction of the bosonic field $A^{2}$ term is
taken into account is a long-standing issue and remains  very
controversial to date\cite{Buzek,KRZ1,KRZ}.

It is known that quantum critical phenomena exhibits deep relations
to the Berry phase (BP) \cite{berry,Guridi,Carollo,Plastina} and the
fidelity \cite{Quan,You,Cozzini,Min,Gu,chens,zhou,Liu}. The BP has
been extensively studied by the geometric time evolution of a
quantum system, providing means to detect the quantum effects and
critical behavior, such as quantum jumps and collapse
\cite{Dusuel,Caro,Zhu}. A  recent proposal is to use the fidelity in
identifying the quantum phase transition\cite{You,Cozzini}. As a
consequence of the dramatic changes in the structure of the ground
states, the fidelity  should drop at critical points. In our
opinion, the quantum-phase-like transitions might also be studied in
terms of the BP and the GS fidelity.

In this paper, we  calculate the ground state BP and fidelity in the
RWA DM with and without $A^{2}$ term to  quantify phenomena of the
quantum-phase-like transitions  in finite system. Without $A^{2}$, a
exact solution to the DM is given explicitly.  We solve the RWA DM
with an additional $A^{2}$ term by a exact diagonalization in the
Fock space of the bosonic operators. The paper is organized as
follows. In Sec.II, we review the RWA DM and the Hamiltonian with
the $A^{2}$ term to obtain exact solutions respectively. In Sec.III,
we study the instability of the RWA DM by measuring the BP and the
ground state fidelity.   The behaviors  of these two quantities as a
function of the interaction strength of the field for the RWA DM
with $A^{2}$ term are also evaluated. Finally, we present the
conclusion in Sec.IV.

\section{Model}

\subsection{Exact solution to the RWA DM}

Let us consider DM of $N$ two-level atoms with energy level $\omega
_0$, interacting with a single-mode bosonic field with the frequency
$\omega $. In the RWA DM, ignoring the counter-rotating term, the
corresponding Hamiltonian is given by
\begin{equation}\label{H2}
H=\omega a^{\dagger}a+\omega _0J_z+\frac \lambda
{\sqrt{N}}(a^{\dagger}J_{-}+aJ_{+}).
\end{equation}
where  $a^{\dagger}$ and $a$ are the photonic creation and
annihilation operators, $  J_k(k=z,\pm )$ denotes the collective
spin-$1/2$ atomic operators, $\lambda $ is the atom-field coupling
strength, and $\hbar \;$is set unity.

Motivated by the exact technique of the  Jaynes-Cammings model with
RWA, we present a detailed  numerical diagonalization procedures to
solve a set of closed equations to the DM with RWA, which was also
briefly discussed in Ref. \cite{Buzek}. Since the Hamiltonian (
\ref{H2}) commutes with the total excitation number operator
$\hat{L}=a^{\dagger}a+J_z+\frac{1}{2}$, the subspace of the Hilbert
space consists of a sum of subspaces labeled by different number of
excitations $L$. The Hilbert space of the collective algebra is
spanned by the kets $\{|j,m\rangle; m=-j,-j+1,\cdots,j\}$. By
adapting Schwinger's representation of spin in terms of harmonic
oscillators \cite{schwinger,sakurai}, $|j,m\rangle$ can be expressed
as $|j,N-n\rangle$, which is a Dike state of $N-n$ spin-up atoms and
$n$ spin-down atoms, $n=0,1,...N$. In this work, $j$ takes its
maximal value $N/2$. $|N/2,N-n\rangle$ is also known as the
eigenstates of $J_{z}$ and $J^{2}$ with
$J_{z}|\frac{N}{2},N-n\rangle=(\frac{N}{2}-n)|\frac{N}{2},N-n\rangle$.
The action of the corresponding raising and lowering operators on
this state gives
$$
J_{+}|\frac{N}{2},N-n\rangle=\sqrt{(N-n+1)n}|\frac{N}{2},N-n+1\rangle
$$
$$J_{-}|\frac{N}{2},N-n\rangle=\sqrt{(N-n)(n+1)}|\frac{N}{2},N-n-1\rangle$$.

\begin{figure}[tbp]
\includegraphics[width=8cm]{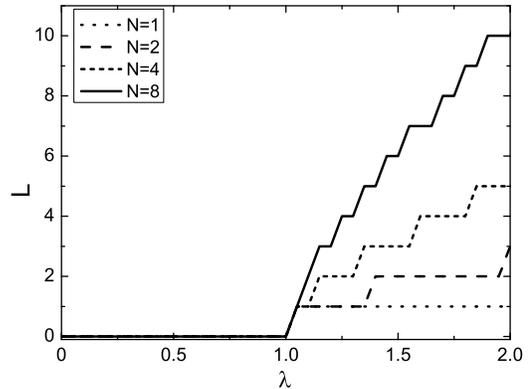}
\caption{Excitation number $L$ versus atom-field coupling parameter
$\lambda$ for different numbers of atoms $N=1,2,4,8$.}
\label{exitation}
\end{figure}

In the subspace of $L=N+k$ excitations the wave function is supposed
as
\begin{eqnarray}\label{wave function}
\left| \psi \right\rangle&=&\sum_{n=0}^Nc_n\left|
n+k\right\rangle_{f}
\bigotimes\left| N/2,N-n\right\rangle_{a}  \nonumber \\
(k&=&-N,-N+1,...)
\end{eqnarray}
where $c_{n}$'s are coefficients, $|n+k\rangle_{f}$ is a Fock state
of the bosonic field with an alterative number $k$, ranging from
$-N$ to infinity. $k$ is equal to $-N$ in the weak coupling regions
corresponding to $0$ excitations and then increases with the
coupling parameter $\lambda$. When the number of excitations $L$ is
larger than the number of atoms $N$, i.e. $k>0$, the ground state of
$H$  lies in the subspace spanned by $N+1$ vectors. In this way, the
Hamiltonian is expressed by a tridiagonal $(N+k+1)\times(N+k+1)$
matrix. From Eq.( \ref{wave function}) we obtain the exact
expression of the $m$-th row of the Schr\"{o}dinger equation
\begin{eqnarray}\label{eq1}
Ec_m &=&\frac \lambda {\sqrt{N}}\sqrt{m+k}\sqrt{(N-m+1)m}c_{m-1}  \nonumber \\
&&+\frac \lambda {\sqrt{N}}\sqrt{m+k+1}\sqrt{(N-m)(m+1)}c_{m+1}  \nonumber \\
&&+[\omega (m+k)+\omega _0(N/2-m)]c_m
\end{eqnarray}
where \begin{equation}m=\left\{\begin{array}{r@{\quad:\quad}l}
 $0,1,...N+k$  &  k\leq 0   \\ $0,1,...N$  &  k>0 \end{array}\right.
\end{equation}
Note that the above equations are closed and the set of linear
equations for $c^{\prime }$s takes a tridiagonal form. Solutions for
a given $k$ are readily obtained through Gaussian elimination and
back substitution. Finally the chosen $k$ corresponds to the lowest
energy among eigenvalues of the solutions for a fixed coupling
strength $\lambda$. It is interesting to find that the excitation
number $L$ is added step by step and keeps a constant in a coupling
parameter interval $[\lambda_{i},\lambda_{j}]$, where
$\lambda_{i}(\lambda_{j})$ is a quantum-phase-like transition point,
as shown in Fig. \ref{exitation}. The first transition point is
denoted as  $\lambda_{c}^{0}$.  The sensitive quantities like  the
ground state BP and the fidelity will be calculates to quantify the
discontinuities, so called instability, in the finite DM with RWA.
\subsection{Numerical exact diagonalization to the RWA DM with $A^2$ term}

It is interesting to discuss the effect of the interacting bosonic
field in the atom-field system. As the interactions vector potential
$A$, caused by the longitudinal part of the bosonic field, are taken
into account, the Hamiltonian in the RWA DM can be evaluated with an
additional term $A^{2}$ \cite{KRZ1,KRZ2}.The additional term $
A^{2}$   has been discussed classically about thermodynamic
properties by Rz\c{a}\.{z}ewski et al \cite{KRZ2,KRZ3}. To
extensively quantify the contributions of the $A^{2}$ term, we
employ  the quantum information tools such as  the BP and the ground
state fidelity to  detect the quantum-phase-like transitions.

In terms of the  bosonic operators, the $A^{2}$ term is given by
$\varepsilon(a^{\dagger}+a)^{2}$, where $\varepsilon$ is the
interacting strength of the bosonic field. The overall Hamiltonian
of the ensemble of two-level atoms interacting with the bosonic
field is expressed as
\begin{equation}\label{A^2}
H_{A}=\omega a^{\dagger}a+\omega _0 J_{z}+\frac \lambda {\sqrt{N}%
}(a^{\dagger}J_{-}+aJ_{+})]+\varepsilon (a^{\dagger}+a)^{2}
\end{equation}
In order to obtain the numerical exact solution, we perform a
standard Bogoliubov transformation by introducing bosonic
annihilation (creation) operator $b(b^{\dagger})$, such that $
b^{\dagger}=\mu a+\nu a^{\dagger}$ and $|\mu|^{2}-|\nu|^{2}=1$.
After substituting  $a$, $a^{\dagger}$ into Eq.( \ref{A^2}) the
total Hamiltonian is diagonalized as
\begin{eqnarray}\label{HA}
H_{A}&=&\sqrt{\omega^{2}+4\omega\varepsilon} b^{\dagger}b+\omega
_0
J_{z}+\frac{1}{2}(\sqrt{\omega^{2}+4\omega\varepsilon}-\omega)\nonumber \\
&+&\frac{\lambda}{\sqrt{N}}[\mu(b^{\dagger}J_{-}+bJ_{+})-\nu(b^{\dagger}J_{+}+bJ_{-})]
\end{eqnarray}
where
$$
\mu^{2}=\frac{1}{2}(\frac{\omega+2\varepsilon}{\sqrt{\omega^{2}+4\omega\varepsilon}}+1),
\nu^{2}=\frac{1}{2}(\frac{\omega+2\varepsilon}{\sqrt{\omega^{2}+4\omega\varepsilon}}-1)
$$
Note that a counter-rotating term is included  in the modified
Hamiltonian ( \ref{HA}), which may plays an essential role in the
following discussion. Because the $A^{2}$ term breaks the gauge
invariance of the Hamiltonian( \ref{H2}) in the DM with RWA, it was
argued in Ref. \cite{KRZ} the instability would be then eliminated.

We now consider the wave functions of the total Hamiltonian with
$N$ atoms, which are of the form
\begin{eqnarray}
|\varphi\rangle_{A}=\sum_{n=0}^{N}\sum_{m=0}^{Ntr}d_{nm}|m\rangle_{f}|N/2,n\rangle_{a}
\end{eqnarray}
where $Ntr$ is the maximum photonic number in the artificially
truncated Fock space, and $d_{nm}$ are coefficients. $|m\rangle_{f}$
is a Fock state with $m$ photons. $|N/2,n\rangle_{a}$ is a Dicke
state in Schwinger's representation of spin with $n$ atoms in
excited state. The $m$-th row of the Schr\"{o}dinger equation reads
\begin{eqnarray}\label{eq2}
Ed_{nm}&=&[\omega_{\varepsilon}m+\Delta(n-\frac{N}{2})+\frac{1}{2}(\omega_{\varepsilon}-\omega)]d_{nm}\nonumber \\
&+&\frac{\lambda\mu}{\sqrt{N}}\sqrt{(m+1)(N-n+1)n}d_{n-1,m+1}\nonumber \\
&+&\frac{\lambda\mu}{\sqrt{N}}\sqrt{m(n+1)(N-n)}d_{n+1,m-1}\nonumber \\
&-&\frac{\lambda\nu}{\sqrt{N}}\sqrt{(m+1)(n+1)(N-n)}d_{n+1,m+1}\nonumber \\
&-&\frac{\lambda\nu}{\sqrt{N}}\sqrt{m(N-n+1)n}d_{n-1,m-1}
\end{eqnarray}
The eigenvalues and eigenfunctions can be obtained numerically by
diagonalizing a $(N+1)\times(Ntr+1)$ matrix. The BP and the fidelity
can be calculated through these eigenfunctions.

To be complete, we also briefly review the contribution of the
$A^{2}$ term in the DM model without RWA, which yields some
unimportant corrections. The Hamiltonian of the full  DM with the
$A^{2}$ term reads
\begin{equation}
H_{DM}=\omega a^{\dagger}a+\omega _0 J_{z}+\frac \lambda {\sqrt{N}%
}(a^{\dagger}+a)J_{x}+\varepsilon (a^{\dagger}+a)^{2}
\end{equation}
With a rotation around an $y$ axis by an angle $\frac{\pi}{2}$ and
the same Bogoliubov transformation, the modified Hamiltonian
$H_{DM}$ is rewritten as
\begin{eqnarray}\label{DM}
H_{DM}&=&\sqrt{\omega^{2}+4\omega\varepsilon} b^{\dagger}b-\omega
_0
J_{x}+\frac{1}{2}(\sqrt{\omega^{2}+4\omega\varepsilon}-\omega)\nonumber \\
&+&\frac{2\lambda}{\sqrt{N}}(\mu-\nu)(b^{\dagger}+b)J_{z}.
\end{eqnarray}
Therefore, $A^{2}$ term in the DM model without RWA would not change
the nature of the phase  transition, except that the position of the
critical point is shifted.

\section{Ground state property}
\subsection{Instability in the RWA DM}

\begin{figure}[tbp]
\includegraphics[width=8cm]{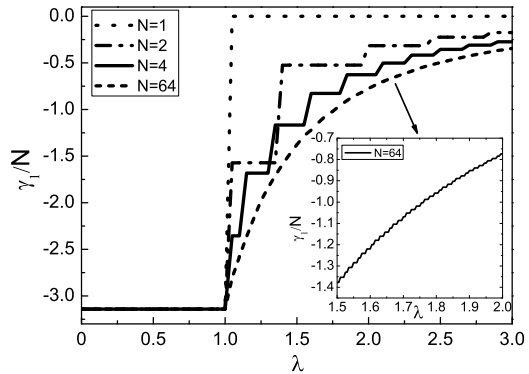}
\caption{ The average Berry phase $\gamma_{1}/N$ of the RWA DM as a function of the coupling parameter $%
\lambda$ for different numbers of atoms $N=1,2,4,64$. The inset
shows a discontinuous picture of instability for $64$ atoms. }
\label{berry phase}
\end{figure}
Berry's adiabatic geometric phase describes a phase factor of the
wave functions in a time-dependent quantum system. The interesting
paths of evolution for generating a BP are those for which the
ground state of the system can evolve around a region of
criticality. We first  measure a nontrivial BP circulating  a region
of "criticality" corresponding to a abrupt change. The BP
$\gamma_{1}$ generated after the system undergoing the
time-dependent unitary transformation $U(T)=\exp[-i\phi(t)J_{z}]$,
varying the angle $\phi(t)$ adiabatically from $0$ to $2\pi$, can be
evaluated as a function of the atom-field coupling parameter
$\lambda$
\begin{equation}  \label{BP1}
\gamma_{1}=i\int^{2\pi}_{0}\langle
\psi_{0}|U^{\dagger}(t)\frac{d}{d\phi}U(t)|\psi_{0}\rangle d\phi
=2\pi\langle \psi_{0}|J_{z}|\psi_{0}\rangle
\end{equation}
where $|\psi_{0}\rangle$ is the ground-state wave function of
Hamiltonian (\ref{H2}) of the RWA DM. As shown in Fig. \ref{berry
phase}, the first phase-transition-like occurs at the "critical"
value of the coupling parameter $\lambda_{c}^{0}=1$ for arbitrary
atom number $N$, which recover the result in the thermodynamical
limit \cite{Gang}. The   average BP $\gamma_{1}/N$ is equal to $\pi$
at $\lambda<1$ and increases abruptly at discontinuous "critical"
points when $\lambda>1$. Note that the plateau is formed  clearly
for $N=1,2,4$ , the width of the plateau becomes narrower and
narrower with the increasing $N$, which are quite different from the
phenomenon of the quantum phase transition in the full  DM
\cite{Lieb,Emary,Chen}. A clear picture of the instability in the
ground state of the RWA DM is given in terms of the BP with $N=64$
atoms shown in the inset of Fig. \ref{berry phase}.

The effect of decoherence of the driving field on adiabatic
evolutions of spin and quantized modes system has been investigated
\cite{Guridi,Carollo}. In the fully quantized context we need a
procedure capable of generating an analogous phase change in the
state of the field. The BP $\gamma_{2}$ is obtained in terms of the
bosonic operator by the phase shift unitary operator
$R(\phi)=\exp[-i\phi(t)\hat{n}]$, where $\hat{n}=a^{\dagger}a$ is
the number of bosons in the field. Changing the angle $\phi(t)$
slowly from $0$ to $2\pi$ the ground state $\gamma_{2}$ is given by
\begin{equation}  \label{BP2}
\gamma_{2}=i\int^{2\pi}_{0}\langle
\psi_{0}|R^{\dagger}(t)\frac{d}{d\phi}R(t)|\psi_{0}\rangle d\phi
=2\pi\langle \psi_{0}|a^{\dagger}a|\psi_{0}\rangle
\end{equation}

\begin{figure}[tbp]
\includegraphics[width=8cm]{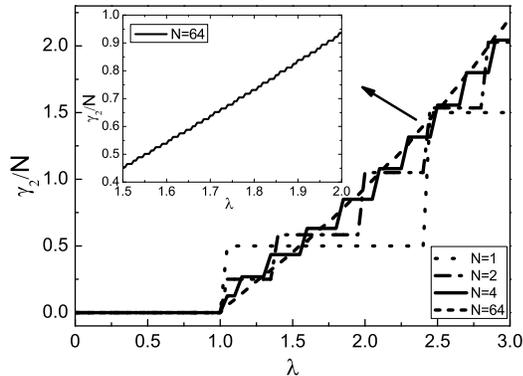}
\caption{The average Berry phase $\gamma_{2}/N$.i.e.  the average
phonon number, in units of $2\pi$ of the RWA DM as a function of the
coupling parameter  $\lambda$ for different number of atoms $N$. The
inset gives a discontinuous picture of instability for $64$ atoms.
\label{phonon}}
\end{figure}

We now have a general expression for the BP $\gamma_{2}$ related to
the photonic number,which is  driven by fields. We plot behaviors of
$\gamma_{2}/N$ in units of $2\pi$ as a function of the atom-field
coupling parameter $\lambda$ for different number of atoms $N$ in
Fig. \ref{phonon}. The BP $\gamma_{2}/N$ is  $0$ in the weak
coupling region for $\lambda\leq 1$ and then increases
discontinuously as $\lambda$ increases. As shown in Fig.
 \ref{phonon}, when $N$ increases the interval of the ``critical"
values $\lambda$ become smaller, leading to the curve with more
steps. The inset of Fig. \ref{phonon} shows that there actually
exist many phase-transition-like ``critical" points beyond
$\lambda=1$ for large $N=64$.

\begin{figure}[tbp]
\includegraphics[width=8cm]{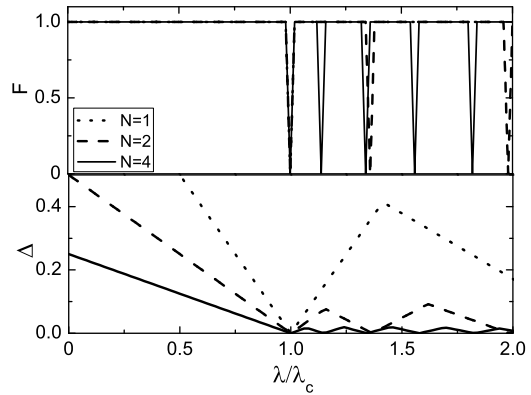}
\caption{Ground state fidelity $F$ and energy gap $\Delta$ between
the energies of the first excited  and ground state in the RWA DM as
a function of the coupling parameter of $\lambda/\lambda_{c}^{0}$
for different number of atoms $N=1,2,4,\infty$.} \label{fgap}
\end{figure}

An increasing interest has been drawn in the role of the ground
state fidelity in detecting the quantum phase transitions for
various many-body system, with a narrow drop at the transition
point. Below we propose to use  this quantum tool to identify the
quantum-phase-like transitions, where the GS fidelity drops to $0$
in the RWA DM. It is defined as the overlap between two ground
states $|\psi_{0}(\lambda)\rangle$ and
$|\psi_{0}(\lambda+\delta\lambda)\rangle$ \cite{Min,Gu}, where
$\delta\lambda$ is a tiny perturbation parameter, i.e.
\begin{equation}  \label{fid}
F(\lambda,\delta\lambda)=|\langle\psi_{0}(\lambda)|\psi_{0}(\lambda+\delta%
\lambda)\rangle|
\end{equation}
Note that $F$ is a function of both $\lambda$ and $\delta\lambda$.
Based on the normalized and orthogonalized wave function in
Eq.(\ref{wave function}) for the RWA DM, the ground state fidelity
can be derived analytically
$|\sum_{n,m=0}^{N}c_{n}(\lambda)c_{m}(\lambda+\delta\lambda)
\delta_{n,m}\delta_{n+k,m+k^{'}}|$, and then can be simplified as
\begin{equation}\label{W2} F(\lambda,\delta\lambda)=
\left\{\begin{array}{r@{\quad:\quad}l}
 $0$  &  n=m,k\neq k^{'}   \\ $1$  &  n=m,k=k^{'} \end{array}\right.
\end{equation}
At each transition point in RWA DM, the  alternative number $k$ is
changed abruptly  and $F$ is then  equal to $0$. Beyond "critical"
points,  $F$  should  be constant $1$.  The energy gap $\Delta$
between the first excited and ground state energies tends to $0$ at
each "critical" point, as shown in Fig. \ref{fgap}. We attribute
this level crossing to the fact that the GS wave functions at the
different sides of each transition point are orthogonal. The ground
state fidelity drops to $0$ at each  "critical" point, exhibiting
similar critical singularity of the first-order quantum phase
transition \cite{chens}. It is also obvious that the instabilities
increase with a increasing number of atoms.

\subsection{Behaviors of the Berry phase and fidelity in the RWA DM with $A^{2}$ term }

\begin{figure}[tbp]
\includegraphics[width=7cm]{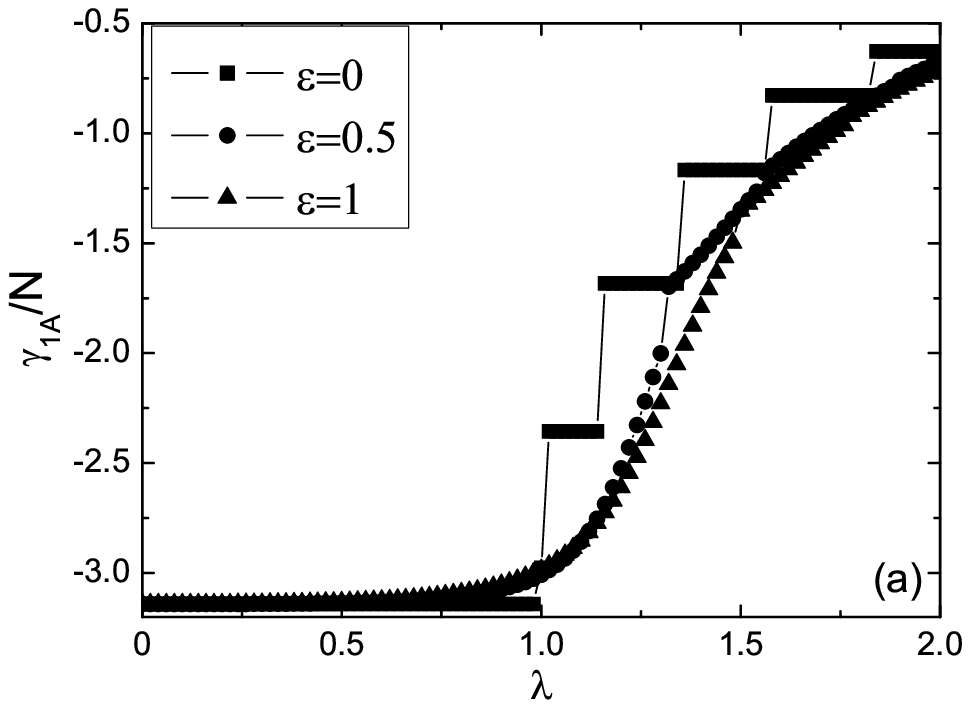}
\includegraphics[width=7cm]{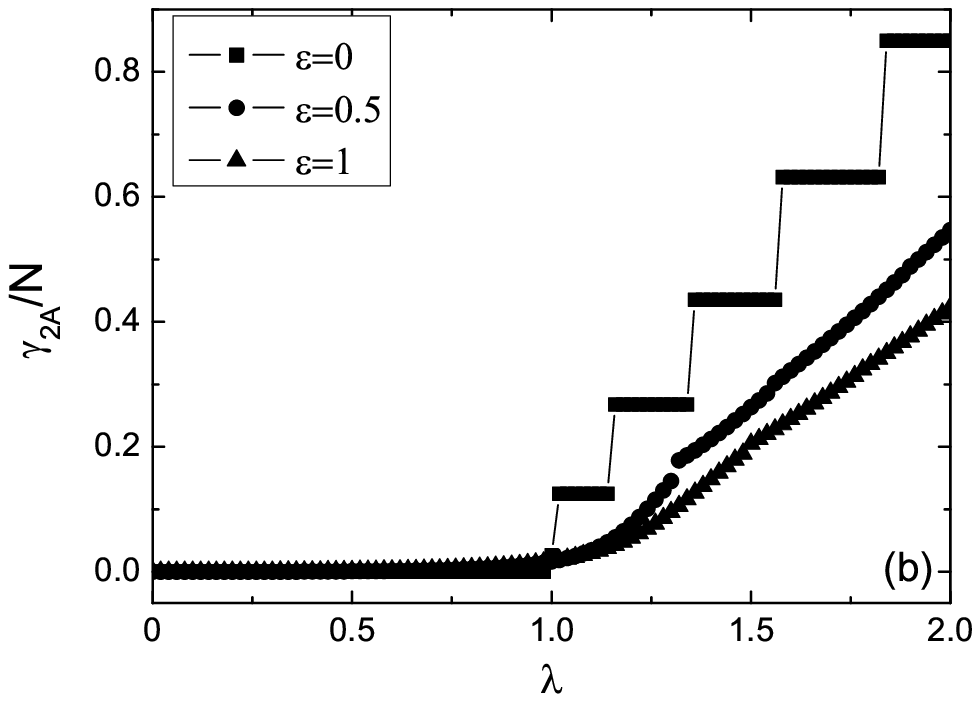}
\includegraphics[width=7cm]{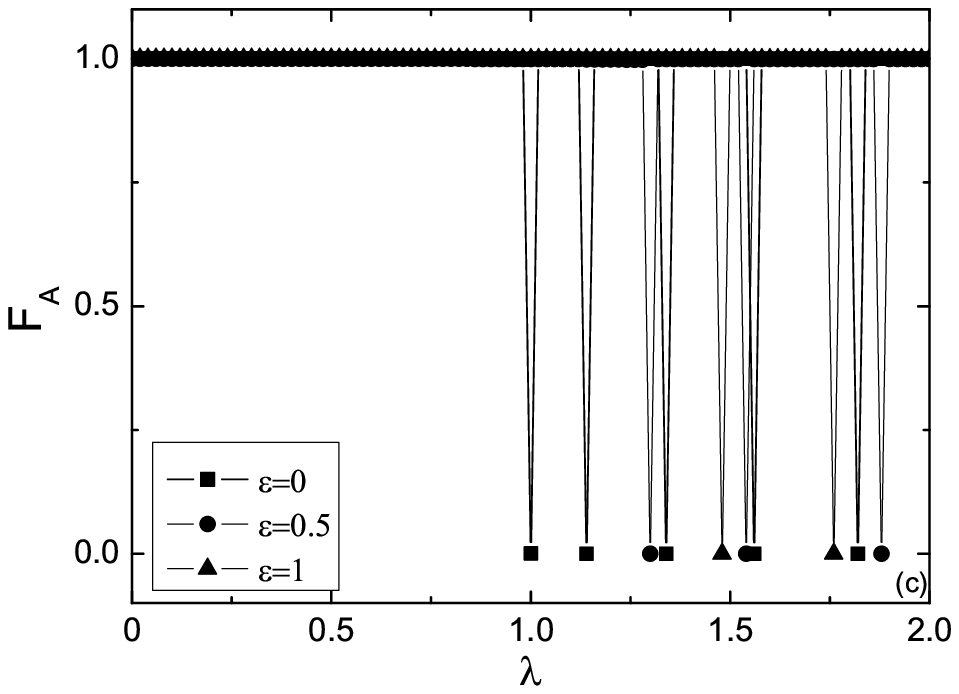}
\caption{\label{N4}  The average Berry phase (a) $\gamma_{1A}/N$,
(b) $\gamma_{2A}/N$ in units of $2\pi$,  as well as  (c)  ground
state fidelity $F_{A}$ of the RWA DM with $A^{2}$ term versus
$\lambda$ for different interacting strengthes of the field
$\varepsilon=0,0.5,1$ for $N=4$. }
\end{figure}

\begin{figure}[tbp]
\includegraphics[width=8cm]{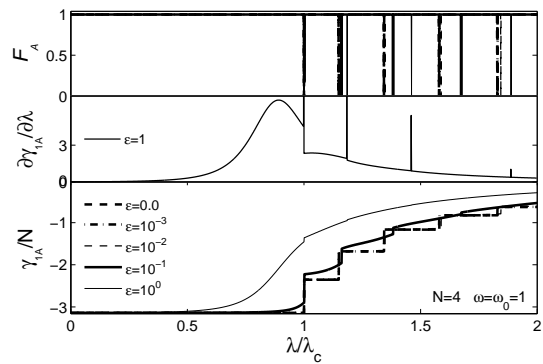}
\caption{\label{n4fgap} Ground state fidelity $F_{A}$ (a), Berry
phase $\gamma_{1A}/N$ (b) and its first derivative $\partial
\gamma_{1A}/\partial\lambda$ (c) of the RWA DM with $A^{2}$ term as
a function of the coupling of $\lambda/\lambda_{c}$ for different
interacting strengthes of the field
$\varepsilon=0,10^{-3},10^{-2},10^{-1},1$. }
\end{figure}
We next turn to study the RWA DM with $A^{2}$ term  by using the
above quantum information tools. By manse of the general BP formula
Eqs.( \ref{BP1}) and ( \ref{BP2}),  the ground state BP of the RWA
DM with $A^{2}$ term can be evaluated as a function of the
atom-field coupling $\lambda$ and the interaction strength of the
field $\varepsilon$, i. e.,
\begin{eqnarray}
\gamma_{1A}=2\pi\langle \varphi_{A}|J_{z}|\varphi_{A}\rangle,\nonumber \\
\gamma_{2A}=2\pi\langle \varphi_{A}|
b^{\dagger}b|\varphi_{A}\rangle
\end{eqnarray}
where $|\varphi_{A}\rangle$ is the ground state of the Hamiltonian
(\ref{A^2}).

We plot the average BP   $\gamma_{1A}/N$ in Fig. \ref{N4}(a) and
$\gamma_{2A}/N$ in units of $2\pi$ in Fig. \ref{N4}(b) as a function
of  the coupling $\lambda$ for different interaction strengthes
$\varepsilon=0,0.5,1$ for $N=4$ atoms. We can observe  that, for
both $\gamma_{1A}/N$ and $\gamma_{2A}/N$, the abrupt jump occurs at
the same coupling parameter $\lambda$ for the same interaction
strengthes $\varepsilon$. We denote the coupling parameter $\lambda$
where the first abrupt jump occurs as a characteristic parameter
$\lambda_{c}$.  It is interesting to observe that the characteristic
$\lambda_{c}$ increases  with the interaction strength
$\varepsilon$. The   stability claimed in Ref. \cite{KRZ} is only
found in the smooth curves for $\lambda\leq\lambda_{c}$. Thus, there
still exist  quantum-phase-like transitions when the interaction of
the bosonic field $\varepsilon$ is strong.

According to the ground state wave function $|\varphi_{A}\rangle$,
the fidelity of the RWA DM with $A^{2}$ term can be also calculated
$F_{A}(\lambda,\delta\lambda)=|\langle
\varphi_{A}(\lambda)|\varphi_{A}(\lambda+\delta\lambda)\rangle|$.
The numerical results   for the different $\varepsilon$ with $N=4$
atoms are exhibited in Fig. \ref{N4}(c). The singularities at the
"critical" points for  $\varepsilon=0,0.5,1$ are demonstrated by a
sudden drop of $F_{A}$. It is clearly shown that  the characteristic
$\lambda_{c}$  moves towards the right regime with the increasing
$\varepsilon$, providing the evidence of the quantum-phase-like
transitions even for a strong interaction of the field.

To show the instabilities more obviously, for a more wide range of
interacting strengthes $\varepsilon=0,10^{-3},10^{-2},10^{-1},1$ ,
we plot the ground state fidelity $F_{A}$, Berry phase
$\gamma_{1A}/N$ and its first derivative
$\partial\gamma_{1A}/\partial\lambda$ of the RWA DM with $A^{2}$
term as a function of the scaled coupling parameter
$\lambda/\lambda_{c}$ together in Fig. \ref{n4fgap}. When the
interaction strength of the field increases the fidelity $F$ still
drops to $0$ at the characteristic $\lambda_{c}$, where the first
derivative $N^{-1}\partial\gamma_{1A}/\partial\lambda$ also changes
abruptly. This is another piece of evidence that the instability of
the RWA DM does not vanish in the RWA DM including the $A^{2}$ term.

It is illustrated that the contribution of the  $A^{2}$ term does
not eliminates the instability of the RWA DM, contrarily to the
previous studies by Rz\c{a}\.{z}ewski et al.\cite{KRZ}. For strong
interaction strength of the bosonic field $\varepsilon$, a
characteristic parameter $\lambda_{c}$ becomes   larger than
$\lambda_{c}^{0}=1$ in the absence of $A^{2}$ term.  A sequence of
the ground state stability reported previously only appears for
$\lambda\leq\lambda_{c}$.

\section{Conclusions}
In summary, we have investigated the instability of the RWA DM by
quantum information tools such as  the BP and the ground state
fidelity. An obvious discontinuous behaviors of these quantities
with finite $N$ atoms are observed. It is demonstrated that the
quantum-phase-like transitions occur beyond the characteristic
$\lambda_{c}$ for strong interaction of fields. We propose that the
instability would not be eliminated by involving the $A^{2}$ term of
the DM with RWA. Previous observed instability may be limited to the
coupling regime $\lambda\leq\lambda_{c}$. It should be pointed out
that the quantum information tools are  very sensitive quantities to
detect quantum phase (or like) transition.

\section{Acknowledgements}
This work was supported by National Natural Science Foundation of
China, PCSIRT (Grant No. IRT0754) in University in China, National
Basic Research Program of China (Grant No. 2009CB929104), Zhejiang
Provincial Natural Science Foundation under Grant No. Z7080203, and
Program for Innovative Research  Team  in Zhejiang Normal
University.

$^{\dag}$ Corresponding author. Email:qhchen@zju.edu.cn

\end{document}